# Play With Me?
# Understanding and Measuring the Social Aspect of Casual Gaming

**Adam Alsén**
Wooga, Analytics and
Data Science
adam.alsen@wooga.com

**Julian Runge**
Wooga and Humboldt
University, Berlin
julian.runge@wooga.com

**Anders Drachen**
Aalborg University and
The Pagonis Network
andersdrachen@gmail.com

**Daniel Klapper**
Humboldt University
Berlin
klapper@hu-berlin.de

**Abstract**

Social gaming is today a pervasive phenomenon. Driven by the advent of social networks and the digitization of game distribution. In this paper the impact of digitization and social networks such as *Facebook* on digital games is described and evaluated. This impact follows several vectors, including the introduction of new game formats and extending the traditional audiences for games, which in turn has increased industrial revenue. The industry is in turn shaped by new business model such as Free-to-Play, digital distribution and the use of viral social features. These changes do not only appear irreversible, but more importantly, play a part in shaping the future of digital game design, notably for mobile devices. The paper presents new knowledge from controlled live experiments from a casual social game across *Facebook* and mobile platforms, finding positive returns by adding social gameplay features. This suggests that not only social network games, but also casual mobile games can benefit from deeper social gameplay mechanics. Given the impact of social features on gameplay, Game Analytics will need to evolve to be able to handle requirements that arise from the introduction of social features, e.g. how to measure engagement against social features and shaping organic and viral spreading of a game.

## I. Introduction

Within the past decade, the proliferation of social networks, notably *Facebook,* led to the emergence of a new type of game, Social Network Games (SNGs) [1-2]. SNGs are played via online social networks [3], make use of the social features offered by the network and extend them with their own social functionalities and asynchronous multiplayer mechanics. In the case of *Facebook*, players can send notifications to friends that invite them to play along and offer them free virtual goods, they can post to friends' walls and they can share status updates from the game to the news feed.

Additionally, some of the games allow for in-game social interactions like visiting a friend's garden in *Farmville* or competing with friends in tournaments or battlefields. SNGs on *Facebook*, and Social Gaming with it, experienced massive viral growth pretty much simultaneously with the platform itself. SNGs are a pervasive phenomenon: By August 2012, *Facebook* had 235 million monthly active players which amounts to roughly a fourth of its monthly active user base at that time [4].

To enable viral spread of the games on the social network, developers made these games very accessible through simple gameplay and well-designed tutorials aiming at onboarding players smoothly [5-6] as well as by using the and by using the Free-to-Play (F2P) business model [7]. This high accessibility and social mechanics made the games appeal to audiences that were not previously involved with gaming.

In parallel with the rise of SNGs, the introduction of mobile platforms enabled games to spread from social networks and unto this new platform. This led to new game forms focusing on the mobile platform, of which many inherited the social mechanics of the SNGs, including linking with the *Facebook* accounts of the players. While *Facebook* is just one of many online social networks (e.g. Steam, the Playstation Network and the Xbox Live service), it is by far the most dominant social network for mobile games, as is evidenced by the *Facebook* account connect offered in the 100 top grossing mobile titles. While most mobile games have social features, they do not always work equally well and questions such as when to include *Facebook*-connect, when to use deeper social gameplay, and when no, are pressing issues in the mobile game industry.

Some casual mobile games are played directly via a hosting social network similar to SNGs, whereas others simply integrate some of the same social mechanics. However, social mechanics are not common in casual mobile games, despite the success of some titles that take advantage of social gameplay or link to social networks such as *Facebook*, e.g. *Clash of Clans* and *Candy Crush Saga* [24]. In summary, despite the popularity of SNGs and social casual mobile games, there has been no work published which investigate the effect of introducing social networks in mobile games. This is in contrast to work focusing on the impact of social networks in businesses and industries outside games [3,8-9].

## II. Contribution

This paper has three main contributions: 1) We present a condensed review of the impact that social networks have had on gaming and the games industry, across SNGs and casual social mobile games. Due to the lack of academic work on SNGs and mobile social games, knowledge from the industry forms the majority of the current state-of-the-art. 2) The paper presents an analysis and description of the game-session play patterns and audience characteristics of three SNGs. 3) We present evidence from a live, controlled experiment investigating the impact of introducing social gameplay features on an existing mobile game, *Diamond Dash*, showing positive returns via the adding of social gameplay features. This suggests that social mechanics, as found in SNGs, provide a potential source of increased return in casual mobile games. Finally, approaches towards including social features in mobile games are discussed.

## III. Background: Casual Social Games

In this paper the term "social game" is used to describe any game that includes social gameplay features and thus either permit or require social interaction. This covers a broad variety of digital and non-digital games. However, here the focus is specifically on "casual social games" (CSGs), which form a subset of social games, and adheres to four design principles outlined below. The term "casual" here refers to the minor temporal investment needed overall and per session by the players to play these kinds of games. A casual game can be played infrequently and in short sessions, while potentially also enabling more frequent and longer-session play. A casual gamer plays infrequently, and in short sessions. There are no standards defining the limits on the term "casual" nor "hardcore", and a broad definition fits the current investigation well as we are targeting the discussion from a top-down perspective.

The term "social game" can be used as an umbrella concept, which include also SNGs. Sometimes SNG and social games are used interchangeably [10], and SNGs did bring social games much of the widespread distribution this game form is currently experiencing. SNGs is used to describe games that are hosted on a digital social network, and which integrates the social features of the network. Wohn et al. [2] stated that: *"what constitutes a SNG is determined more by the technical aspects of how it is accessed and distributed, not on the genre of the game"*. Additionally, SNGs can be viewed as a sub-category of social games, as are casual social games. However, casual social games (CSGs) can be SNGs as well as be hosted on any other platform for game delivery, including mobile and console platforms. CSGs refer to a broad category of highly accessible online games with social gameplay, featuring social media integration and/or social browser games outside of social networks. There are a number of defining characteristics for casual social games:

- **High accessibility and engagement**: These games have a flat learning curve and a casual gameplay that is easily accessible [1,5] ("easy-in, easy-out" [11-12]). This allows non-gaming customers to get into the gameplay [10-11,13] and often stay engaged for a long time due to smart game design [1,14].
- **Inclusion of viral/social features**: A distinguishing feature of casual social games are the included social features, e.g. gifting of virtual goods, sharing achievements, posting to friend's walls, that reward viral sharing and hence facilitate sociability and viral growth of the games [1-2,4,15].
- **Free-to-play**: Most casual social games follow the freemium business model where the base version is offered to the customer for free and premium upgrades can be purchased for real money. F2P is the formulation of the freemium business model for gaming [7,13]. Most casual social games offer a form of virtual currency that can be purchased for real money [16-17,30].
- **Strong sociability around the game** [15]: Lewis et al. [1] noted for SNGs that: *"social interactions around SNGs are clearly a central aspect of the pleasure of the experience for many players"*. This feature is also found in CSGs, and is enabled by the accessibility of these games and the resulting large communities around them. The social dimension of CSG reaches beyond the social interaction inside the games [2,15,19-20].

Casual social games cover a variety of genres or types of games. However, the main game genres traditionally are resource management and simulation games (e.g. *Farmville*), Social Casino games (e.g. *Texas Hold'em Poker*) and Casual/Arcade games (e.g. *Diamond Dash*). This is of course subject to change over time. Genres that experienced growth more recently are Strategy, Hidden Object and Match-3.

### Players of Casual Social Games

With the introduction of SNGs, the reach and depth of engagement in games increased. Using historical data and forecasting, Borrell Associates [20] predicted that the average daily time spent playing video games per capita will go from 18 minutes in 2008 to 28 minutes in 2018.

The design of early casual social games made it easy for *Facebook* users to get hooked to the games [1]. The built-in viral features require players to engage their online social networks to remove blockers in the games or obtain rewards. This allowed SNGs to spread to audiences that previously were agnostic or skeptical of gaming (Wohn et al. [2] provide insights into the process of viral spread between users of a social network). The social mechanics used by SNGs generated a completely new audience for online games [13], marked by different demographics and play patterns, the casual gamers. Notably, casual gamers are more likely to be female than other game audiences gamers. This is emphasized by survey results published by the Entertainment Software Association [21]; the gender split of video gamers in the United States went from 38% female and 62% male in 2006 to almost 50-50% in 2014.

### IV. Industry Impact of Casual Social Gaming

Shortly after their introduction, some SNGs and other online games operating under the F2P business model generated substantial revenue from sales of virtual goods through in-app purchases [16-17]. This benefited both *Facebook* that usually retains a share of the revenue, and the companies developing and publishing the SNGs. Rather low entry barriers, free viral growth and immediate revenue opportunities attracted entrepreneurs. New companies in the gaming industry emerged [4]. The Social Gaming space started seeing a lot of investment and mergers and acquisition activity: *"Social and casual games took home the bacon [in 2011], making up 57 percent of private investment and 45 percent of M&A [mergers and acquisitions] activity [in the gaming industry]"* [23].

One of the more prominent examples of such a company is Zynga that went public in 2011 at a seven billion USD valuation. In 2012, five of the top ten developers were from Europe: King.com, Peak Games, Rovio, Social Point, and Wooga [4]. SNGs also affected the gambling industry through the advent of Social Casino Games (SCGs) where players invest real money to gamble for virtual rewards. These games' proliferation was facilitated by the fact that they are not subject to the same tough regulation that real money gambling applications are. King.com (later renamed to King) ported SNGs to mobile and achieved tremendous success with the mobile version of its *Candy Crush Saga* that finally led to its multi-billion USD initial public offering in 2014. The largest part of companies that succeeded with SNGs either extended or even shifted their focus on to mobile games that are highly lucrative and growing. Among these companies are, besides King, Zynga that acquired Natural Motion, branching into mobile gaming and Wooga that ported its apps similar to King. As of now, *Clash of Clans* is the most successful mobile game in terms of revenue, generating millions of dollars per day for its creator Supercell [24].

**New Business Models**

Social Gaming has notably seen the introduction and widespread adoption of the F2P business model, and the use of social features for viral growth as new business tactics. The games themselves have also been used for corporate promotion. The freemium business model is not only used in SNGs, but adopted for non-social mobile games and making its way into the traditional gaming space: *"We've seen this model creep into console games with downloadable map packs for Modern Warfare and Halo, and additional content for games like Guitar Hero that have been highly profitable"* [25]. It can be argued that F2P – itself a result of the virtualization or digitization of many goods and the close to zero marginal cost of production and distribution of these goods [7] – would have found its way into traditional gaming anyways. Social Gaming however undoubtedly accelerated this development by its relentless adoption of F2P. Its effect also shows in modifications to console hardware (the "share button") and the inclusion of social networks in the console and PC gaming experience [10].

**Revenue Impact**

The business activity added through casual social games, particularly on browser and mobile platforms, impacted industry revenue substantially. A new revenue stream from digital gaming, now continuing to grow through mobile gaming, emerged. There are cannibalizing effects on traditional gaming companies and consolidation creeps among developers of casual social games. But overall the revenue impact is positive and gaming is expected to keep growing [26]. The growth of Social Gaming at a compound annual growth rate of 15% until 2019 [10] outpaces the growth of the gaming market overall at 5.5% over the same time period [26]. Traditional gaming companies are realizing the significance and future opportunities, e.g. games using augmented reality, of mobile gaming. They are hence investing in the social and mobile gaming space, exemplified by Activision Blizzard's recent acquisition of King for six billion USD. Casual social games hence shaped and shape the revenue of the video games industry, the strategies of gaming companies and the present and future mobile gaming experience.

**Mobile Casual Social Games**

Investigating the top 150 grossing games in the United States App Store, as of April 2016, one can observe that over 90% of casual games on mobile platforms offer *Facebook* connect to their players, while 100% of all CSGs do the same. Additionally, a number of developers have started their own online social networks inside their player base, also allowing players to find new friends through the game versus offering players to invite/play with connections from their existing social network. These observations underline the importance to understand how the social dynamics introduced by adding social mechanics in casual mobile games influence elements like gameplay, engagement, retention and revenue.

A number of studies have looked into the social dynamics of casual social games [e.g. 1-2,11,14-15,28,31]. These generally indicate the relevance of social interaction as a driver of engagement in games, and social games more broadly. However, none of these studies explore the crucial link between social features, engagement, monetization and revenue generation. It is therefore presently difficult to directly evaluate how important social mechanics/gameplay is for revenue and engagement in mobile casual games. Such empirical would work would however assist with addressing several questions, notably: How much can social gameplay features foster engagement of players in mobile games? How is this effect different for different audiences? Are the returns to social gameplay higher or lower than the returns to single-player gameplay?

Shi et al. [29] investigated how much social dynamics and players' past gameplay drive purchase propensity in F2P games. In the empirical part of this study the authors found that purchase propensity is influenced both by formal and informal social dynamics, echoing the conclusions of Sifa et al. [31] who reported that social interaction as the third-most powerful predictor of purchases in their prediction mode. Shi et al. [29] also reported a moderating effect of informal social dynamics on the effect of players' play history on purchase decisions. While these findings are insightful, there is still the question of the causality of these relationships, as well as quantitative statements on the strength of the associations between individual gameplay, social gameplay, engagement and monetization. A natural extension of the work of Shi et al. [29], which is based on time series variations, not controlled experiments, would be to perform controlled experiments with social features using A/B testing [30].

## V. Social Design of 3 CSGs

In this section the focus will be on the social mechanics and player base of the three SNGs covered.

| Social gameplay element | Diamond Dash | Pearl's Peril | Monster World |
|---|---|---|---|
| Requests (ask/gift virtual goods) | Yes | Yes | Yes |
| Share updates from the game | Yes | Yes | Yes |
| Post to friend's wall | Yes | Yes | Yes |
| Visit a friend's garden/island | No | Yes | Yes |
| Friends' bar | Yes | Yes | Yes |
| Multiplayer | Asynchronous* | Asynchronous* | No |

*Table 1. Social gameplay elements in the casual social games used in this study. *Built into the game over time. Walls, garden/islands and bars are social features in the three games that permit players to visit, view the progress of, and often help other players.*

**Social Mechanics**

Here three casual social games are investigated. *Diamond Dash* and *Pearl´s Peril* occur on both mobile and *Facebook*, while *Monster World* is a mobile-only title. All adhere to the four design principles for CSGs outlined above:

- *Diamond Dash,* a casual/arcade game. *Diamond Dash* is a highly casual puzzle game where players play timed levels and can compete with their friends for a high score.
- *Pearl's Peril,* a hidden object game. *Pearl's Peril* is a hidden object game where players search for items in high-quality graphic scenes and have an island that they can and need to decorate to proceed in the game.
- *Monster World,* a resource management/simulation game. *Monster World* finally is similar to Zynga's much famed *Farmville* [1], but a little less conventional in that players can grow Diamond Bushes and Lemonade Trees. Overall, it has the same mechanics where players plant and harvest crops that they sell to customers to get in-game currency that they can use to expand and decorate their garden. There are additional mechanics that were added after launch facilitating e.g. in-game crafting of items using in-game currency.

The games were developed and published by the Berlin-based games company Wooga, who provided in-depth data from the three games. The three games include various social mechanics (features) (table 1). *Monster World* has a more deeply engaging gameplay than *Diamond Dash* (fig. 1) possibly due to more complex gameplay. All contain social features via *Facebook*. Also,, *Pearl's Peril* and *Monster World* allow players to visit each other's island/garden. *Diamond Dash* and *Pearl's Peril* were extended post-launch with asynchronous multiplayer mechanics.

**Player Base**

The player base of the titles used here, across the platforms they are available on (mobile, *Facebook*) match the pattern of casual social games in general: Two-thirds of the audience of *Monster World* and *Pearl's Peril's* are female. Half of the players of *Monster World* are female aged 25 plus; for *Pearl's Peril* this figure is at 35 plus, it hence has a slightly older player base. A fourth of *Pearl's Peril* players are female aged 55 and older. In *Diamond Dash*, the ma-jority of players are also female, but less extreme than in the other two games. In terms of age, its audience is similar in composition to *Monster World*. The average player of the games under study is in line with findings of a survey of social gamers [22]. According to Ingram [22] the average social gamer is a 43-year old female playing on *Facebook*. This contrasts the, formerly common, image of online gamers being young males. While a majority of social gamers entertains casual play patterns, with three or less sessions per week, some players play several sessions a week. *Monster World* appears to be the most deeply engaging game with the thickest tail in fig. 1, followed by *Pearl's Peril*. *Diamond Dash* is characterized by the most casual play pattern which is to be expected for an Arcade-style game. Fig. 1 underlines the casual nature of SNGs with more than half of the players playing only once per week and a session usually lasting only a few minutes.

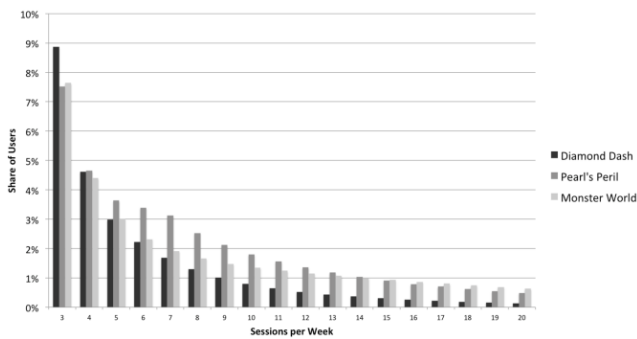

Fig. 1. A histogram of sessions per player per week for the three SNGs evaluated in this paper on *Facebook*; source: Wooga, reprinted with permission.

## VI. Experiment: Impact of Social Features

As noted above, *Diamond Dash* and *Pearl´s Peril* exists as both a *Facebook* (SNG) game as well as a stand-alone mobile titles, but in the latter case without the direct integration of this social network, but instead a connection with the *Facebook* account of the player, which enables e.g. comparison of scores with friends.

Having described the characteristics and impact of CSGs, with an emphasis on the SNG format [8,44], and described the audience of *Monster World, Pearl´s Peril* and *Diamond Dash*, the focus now turns to CSGs on mobile platforms, i.e. games without a direct embedment in a social online network. The motivation for this focus is that current development in CSGs is focused on the very large and rapidly growing mobile platform, while *Facebook* form a much smaller segment. Furthermore, because it is an open question if social features are as important to drive revenue on mobile platforms as on social online networks (results presented here indicate this is the case, see below).

In order to evaluate the impact of introducing social gameplay features in mobile games, an experiment was performed in Wooga´s *Diamond Dash,* using the live game that is available via Apple´s iOS platform as well as on *Facebook*. Such live experiments are extremely rarely reported in academic literature due to their inherently confidential nature [27,30-31]. *Diamond Dash* is F2P which monetizes by offering various in-game currency or power-up purchases. However, the specific monetization mechanics are not the target of the present experiment, but rather the overall impact on revenue across all monetization mechanics, based on adding social mechanics.

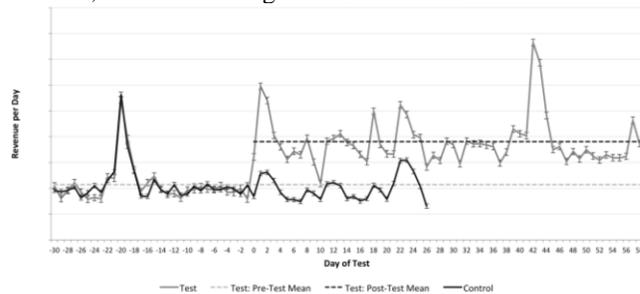

Fig. 2. Team Battles A/B Test in Diamond Dash iOS (mobile platform): Revenue per Day comparison, the test was running from day 0 to 26; source: Wooga, printed with permission.

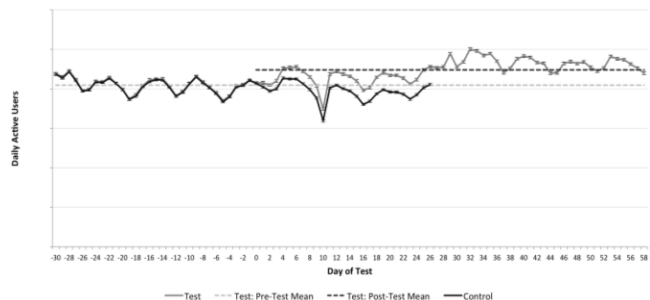

Fig. 3. Team Battles A/B Test in Diamond Dash iOS (mobile platform): Daily active user comparison, the test was running from day 0-26; source: Wooga, printed with permission.

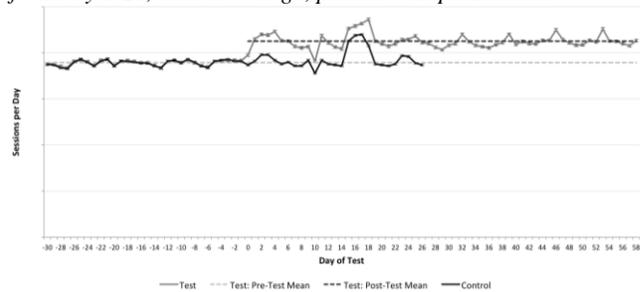

Fig. 4. Team Battles A/B Test in Diamond Dash iOS (mobile platform): Sessions per player per day comparison, the test was running from day 0-26; source: Wooga, printed with permission.

The experiment was designed as an A/B test with pre-test/post-test comparison where a new social feature was introduced to a random subsample of the player base (45% of the player base, randomly selected, for both versions of the game). 45% of the players comprised the control group, the remaining 10% were not included. The total number of players involved in the experiment is over 3 million.

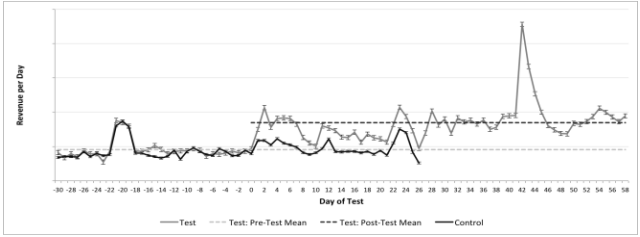

Fig. 5. *Team Battles A/B Test in Diamond Dash iOS (mobile platform): Revenue per Day comparison, the test was running from day 0 to 26; source: Wooga, printed with permission.*

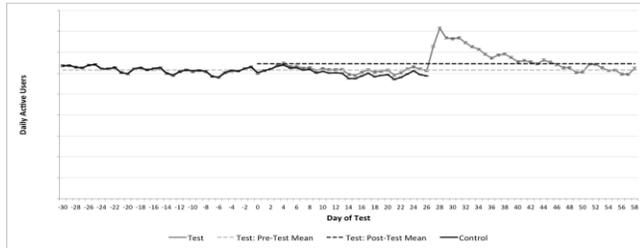

Fig. 6. *Team Battles A/B Test in Diamond Dash iOS (mobile platform): Daily active user comparison, the test was running from day 0 to 26; source: Wooga, printed with permission.*

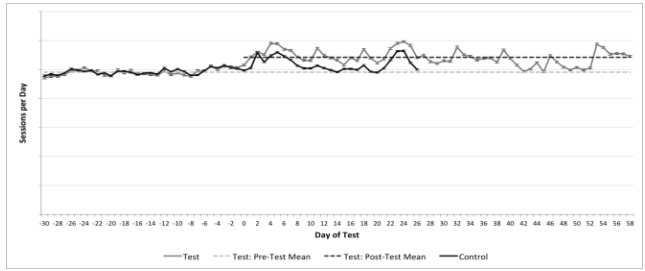

Fig. 7. *Team Battles A/B Test in Diamond Dash iOS (mobile platform): Sessions per player per day comparison, the test was running from day 0 to 26; source: Wooga, printed with permission.*

The social feature, designed for this experiment, was "*Team Battles*" where players battle players in teams. The feature enhances cooperative gameplay inside the teams and competitive between teams. It hence introduces two new social dynamics to the game. No other changes were made to the game while this experiment was running, and thus the introduction of *Team Battles* was the only variable introduced. This is arguably only one example of a social feature in CSGs, but serves as a case example in how social mechanics can potentially impact a CSG.

Figs. 2-4 show the causal effect of this new social feature on monetization (fig. 2), daily active users (fig. 3) and engagement (fig. 4) for the mobile version of *Diamond Dash*. Figs. 5-7 the effect for the *Facebook* version. Please note that the y-axes are not numbered for confidentiality reasons. All three effects are statistically significant and positive at $p<0.01$. More importantly, the result indicated that the effect is persistent, showing not only a short-term uplift but a structural change in monetization and usage. Finally, as can be seen from the charts, the impact on revenue outpaces the impact on daily active users and sessions per day: while we observe an average treatment effect on revenue of close to 90% during the test period, the same number is at 8.5% for daily active users and at roughly 11% for sessions per player per day. At the time, the game had already seen a lot of content being added during its lifetime and yet the relative uplift in revenue compared to any other features added to the game following launch. Another social feature, global tournaments, which introduced a competing league system earlier in the history of the game, also generated revenue uplift. The result indicate that the returns to the additional social gameplay are substantially higher than the returns to the gameplay before the inclusion of the additional social dynamics, and that social gameplay can increase returns in mobile CSGs.

## VII. Discussion and Conclusions

In this paper the impact of online social networks on the games industry has been outlined, showing the emergence of casual social games as an new class of games on new platforms [1-2,13,27], but along with this type of game the coming to life of a new type of player: the casual social gamer [13]. The opportunities created by the growth of this segment has led to the creation of new companies [4]. Casual social games are less characterized by their gameplay or genre [2], but more by their distribution, the inclusion of viral features, the F2P model, accessible gameplay, and the resulting sociability around the games. They are increasingly proliferating to new platforms, namely mobile. This paper presents experimental evidence from live games showing how adding social features increases monetization, engagement and usage across both *Facebook* and mobile platforms. While only one game was used here, results suggest that the returns on social gameplay are higher than on single-player gameplay. If the effect found here is indicative of the advantages of social features across different kinds of mobile games, it follows that Game Analytics, which provides business intelligence across the mobile sector [7,17] should aim to contribute to a better understanding and measurement of the social aspect of casual games. Open questions abound [32] e.g.: What are the effects of adding social features to casual games? What does it actually do to players and gameplay? How should we measure engagement with social features? How do these in turn shape organic and viral spread of a game? These questions deserve attention to build a body of knowledge around the social aspect of CSGs.

**Acknowledgements:** The authors thank Wooga for facilitating the study; Dereck Toker for extremely helpful comments on earlier versions, and Cedrik Neumann for help with data extraction.

# References

[1] C. Lewis, N. Wardrip-Fruin, J. Whitehead, "Motivational game design patterns of 'ville games," Proc. 12th Foundations of Digital Games, FDG 2012, pp. 172–179.

[2] D.Y. Wohn, C. Lampe, R. Wash, N. Ellison, J. Vitak, "The 'S' in Social Network Games: Initiating, Maintaining, and Enhancing Relationships," Proc. 44th Hawaii Int. Conf. on System Sciences, 2011, pp. 1–10.

[3] J. Heidemann, M. Klier, F. Probst, "Online social networks: A survey of a global phenomenon," Computer Networks, 2012, 56(18), pp. 3866–78.

[4] I. Lunden, "Facebook says it now has 235M monthly Gamers, App center hits 150M monthly visitors," in TechCrunch, TechCrunch, 2012. [Online]. Available: http://techcrunch.com/2012/08/14/facebook-says-it-now-has-235m-monthly-gamers-app-center-hits-150m-monthly-users/. [Accessed: Jan 3, 2016].

[5] H. Tyni, O. Sotamaa, and S. Toivonen, "Howdy pardner!," MindTrek '11 Proceedings of the 15th International Academic MindTrek Conference: Envisioning Future Media Environments, vol. 11, pp. 22–29, Sep. 2011

[6] L. Holin, C. Sun, "Cash Trade in Free-to-Play Online Games," Games and Culture, 2011, 6(3), pp. 270–87.

[7] E.B. Seufert, Freemium economics: leveraging analytics and user segmentation to drive revenue. Amsterdam, Boston: Elsevier/Morgan Kaufmann, 2014.

[8] K. Berger, J. Klier, M. Klier, F. Probst, "A Review of Information Systems Research on Online Social Networks," Communications of the Association for Information Systems, 2014, 35(8).

[9] M. Trier, A. Richter, "The deep structure of organizational online networking - an actor-oriented case study: Deep structure of organizational online networking," Information Systems Journal, 2015, 25(5), pp. 465–88.

[10] Research and M. ltd, "Global social gaming market 2015-2019," 2015. [Online]. Available: http://www.researchandmarkets.com/publication/mrvntcu/global_social_gaming_market_20152019. [Accessed: Jan 8, 2016].

[11] D.-H. Shin, Y.-J. Shin. "Why do people play social network games?," Computers in Human Behavior, 2011, 27(2), pp. 852–61.

[12] C. Klimmt, H. Schmid, J. Orthmann. "Exploring the enjoyment of playing browser games," Cyberpsychology and Behavior, 2009, 12(2), pp. 231–234.

[13] T. S. M. Monthly, "How social games changed the games industry," in All, The Social Media Monthly, 2014. [Online]. Available: http://thesocialmediamonthly.com/social-games-changed-games-industry/. [Accessed: Jan 3, 2016].

[14] J. Sung, T. Bjornrud, Y. Lee, D. Y. Wohn, "Social network games: exploring audience traits," Extended Abstracts on Human Factors in Computing Systems,CHI EA '10, 2010, p. 3649.

[15] J. Hamari, A. Järvinen, "Building customer relationship through game mechanics in social games," Business, Technological and Social Dimensions of Computer Games, 2011.

[16] V. Lehdonvirta, "Virtual item sales as a revenue model: identifying attributes that drive purchase decisions," Electronic Commerce Research, 2009, 9(1–2), pp. 97–113.

[17] R. Sifa, F. Hadiji, J. Runge, A. Drachen, K. Kersting, C. Bauckhage, "Predicting Purchase Decisions in Mobile Free-to-Play Games," Proc. AIIDE, 2015.

[18] J. Stenros, J. Paavilainen, and F. Mäyrä, "The many faces of sociability and social play in games," Proc. the 13th International MindTrek, 2009, pp. 82–89.

[19] Y. A. W. de Kort, W. A. IJsselsteijn, and K. Poels, "Digital games as social presence technology: Development of the social presence in gaming questionnaire (SPGQ)," 2007. [Online]. Available: http://repository.tue.nl/663080. [Accessed: Jan 3, 2016].

[20] "Daily time spent playing video games per capita in the United States in 2008, 2013 and 2018," in Statista, 2016. [Online]. Available: http://www.statista.com/statistics/186960/time-spent-with-videogames-in-the-us-since-2002/. [Accessed: Jan 11, 2016].

[21] "Distribution of computer and video gamers in the United States from 2006 to 2016, by gender," in Statista, Statista, 2016. [Online]. Available: http://www.statista.com/statistics/232383/gender-split-of-us-computer-and-video-gamers/. [Accessed: Jan 1, 2016].

[22] Ingram, M., "Average Social Gamer Is a 43-Year-Old Woman", in Gigaom [Online]. Available: http://gigaom.com/2010/02/17/average-social-gamer-is-a-43-year-old-woman/ [Accessed: Jan 8, 2016].

[23] Empson, R., "Led By Social, Gaming Investment & M&A More Than Doubled In 2011; Consolidation Looms", in TechCrunch [Online]. Available: http://techcrunch.com/2012/03/01/global-gaming-investment-report/ [Accessed: Jan 5, 2016].

[24] Grubb, J., "Clash of Clans developer Supercell's revenues tripled in 2014", in GamesBeat [Online]. Available: http://venturebeat.com/2015/03/24/clash-of-clans-developer-supercells-revenues-tripled-in-2014/ [Accessed: Jan 5, 2016].

[25] Silverman, M., "The Influence of Social Gaming on Consoles" in Mashable [Online]. Available: http://mashable.com/2011/02/22/ consoles-social-gaming/#9A5nPegn5Oqr [Accessed: Jan 8, 2016].

[26] Takahashi, D., "U.S. games industry forecast to grow 30 percent to $19.6B by 2019" in Forbes, 2015 [Online]. Available: http://venturebeat.com/2015/06/02/u-s-games-industry-forecast-to-grow-30-to-19-6b-by-2019/ [Accessed: Jan 8, 2016].

[27] J. Runge, P. Gao, F. Garcin and B. Faltings, "Churn prediction for high-value players in casual social games," in Proc. Computational Intelligence in Games, CIG 2014, pp. 1-8.

[28] J. Hamari, J. Koivisto and H. Sarsa, "Does Gamification Work? – A Literature Review of Empirical Studies on Gamification," in Proc. 47th Hawaii International Conference on System Sciences, 2014, pp. 3025-3034.

[29] S. W. Shi, M. Xia, Y. Huang, "From Minnows to Whales: An Empirical Study of Purchase Behavior in Freemium Social Games," International Journal of Electronic Commerce, 2015, 20(2), pp. 177-207.

[30] S. D. Levitt J. A. List, S. Neckermann and D. Nelson (2016): "Quantity discounts on a virtual good: The results of a massive pricing experiment ad King Digital Entertainment". PNAS v113(2), pp. 7323-7328.

[31] Sifa, R.; Hadiji, F., Drachen A. and Runge J., "Predicting purchase decisions in mobile free-to-play games," Proc. Artificial Intelligence in Interactive Digital Entertainment Conference, 2015.

[32] Alsen, A. and Runge, J., "How Social Is Mobile Gaming? Does Facebook Connect really increase engagement?" in Innovation Enterprise [Online]. Available: https://channels.theinnovationenterprise.com/articles/how-social-is-mobile-gaming? [Accessed: Jan 5, 2016].